\documentstyle[preprint,tighten,eqsecnum,aps,floats,psfig,epsfig]{revtex}

\def\ub{{\overline{u}}}
\def\vb{{\overline{v}}}
\def\wb{{\overline{w}}}

\begin{document}
\draft

\title{
Randomly dilute spin models with cubic symmetry.
}
\author{Pasquale Calabrese$\,^1$, Andrea Pelissetto$\,^2$, 
Ettore Vicari$\,^3$ }
\address{$^1$ Scuola Normale Superiore and  I.N.F.N., Piazza dei Cavalieri 7,
 I-56126 Pisa, Italy.}
\address{$^2$ Dipartimento di Fisica dell'Universit\`a di Roma I
and I.N.F.N., I-00185 Roma, Italy.}
\address{$^3$
Dipartimento di Fisica dell'Universit\`a 
and I.N.F.N., 
Via Buonarroti 2, I-56127 Pisa, Italy.
{\bf e-mail: \rm 
{\tt calabres@df.unipi.it},
{\tt Andrea.Pelissetto@roma1.infn.it},
{\tt vicari@df.unipi.it}.
}}

\date{\today}

\maketitle

\begin{abstract}
We study the combined effect of cubic anisotropy and quenched 
uncorrelated impurities on multicomponent spin models. 
For this purpose, we consider the field-theoretical approach
based on the Ginzburg-Landau-Wilson  $\varphi^4$ Hamiltonian
with cubic-symmetric quartic interactions and  quenched
randomness coupled to the local energy density.
We compute the renormalization-group functions to six loops in 
the fixed-dimension ($d=3$) perturbative scheme.
The analysis of such high-order series provides an accurate
description of the renormalization-group flow.
The results are also used to determine the critical behavior
of  three-dimensional antiferromagnetic 
three- and four-state Potts models in the  presence of quenched impurities. 
\end{abstract}


% ========================= BODY =========================
%\narrowtext

\section{Introduction and main results.}

The critical behavior of systems with quenched disorder is of considerable 
theoretical and experimental interest. 
A typical example is obtained by mixing an (anti)-ferromagnetic 
material with a nonmagnetic one, obtaining the so-called randomly dilute 
magnets. They are usually described by using the Heisenberg
Hamiltonian with a random-exchange term
\begin{equation}
{\cal H}_{H,r} = - J \, \sum_{<xy>}  \rho_x \rho_y 
\,\vec{s}_x \cdot \vec{s}_y   , 
\label{lattmodel}
\end{equation}
where $s_{x,i}$ are $M$-component spins and 
$\rho_x$ are uncorrelated random variables, which are equal to one 
with probability $p$ (the spin concentration) and zero with probability $1-p$
(the impurity concentration). The pure system corresponds to $p=1$.
The critical behavior of these systems is well established,
both theoretically and experimentally, 
see, e.g., Refs.
\cite{GL-76,Aharony-76,Stinchcombe-83,PV-r,FHY-01} and references therein.
In particular, a new random Ising 
universality class describes the critical behavior of the random
Ising model (RIM) above the percolation threshold of the magnetic atoms.

The O($M$)-symmetric Hamiltonian (\ref{lattmodel}) is a
rather simplified model for real magnets. In particular,
it does not take into account 
the presence of nonrotationally invariant interactions 
that have only the reduced symmetry of the lattice and that are 
due, e.g., to the spin-lattice coupling and to dipole-dipole interactions.
In this case, a more realistic Hamiltonian is 
\begin{equation}
{\cal H}_{\rm H,c} = - J \, \sum_{<xy>}  
\,\vec{s}_x \cdot \vec{s}_y  
+ a \sum_x \sum_i s_{x,i}^4,
\label{lattmodel-cubic}
\end{equation}
where $a$ is the anisotropy coupling. 
For many materials $a/J$ is relatively small
and thus one usually neglects these additional interactions.
However, this is fully justified only if they are irrelevant in the
renormalization-group (RG) sense. 
This issue may be investigated by considering  the cubic-symmetric
$\varphi^4$ Hamiltonian \cite{Aharony-76}
\begin{equation}
{\cal H}_c = \int d^d x 
\left\{ {1\over 2}(\partial_\mu \phi(x))^2 + {1\over 2} r \phi(x)^2 + 
{1\over 4!} v \left[ \phi(x)^2\right]^2 +
{1\over 4!} w \sum_{i=1}^M \phi_i(x)^4 \right\},
\label{Hphi4cubic}
\end{equation}
where $\phi(x)$ is an $M$-component field and $r\propto T-T_c$.
Analyses of high-order perturbative expansions
(see, e.g., Refs.~\cite{FHY-00,CPV-00,Varnashev-00,SAS-97,KS-95})
show that the cubic-symmetric quartic $w$-interaction is relevant
for $M\geq 3$, and in particular, for the physically relevant case $M=3$.
In this case, the nature of the transition depends
on the sign of the coupling $w$:
if $w>0$, the critical behavior is described by a new fixed
point with reduced cubic symmetry,
while, for $w<0$, the RG flow 
runs away to infinity, and the corresponding system is expected
to undergo a weak first-order transition.	
In the two-component case, the O(2)-symmetric fixed point
is stable with respect to the $w$-perturbation, and thus, if the 
transition is continuous, it belongs to the XY universality class.

It is of interest to study the effect of quenched disorder on 
cubic magnets. As discussed in Ref.~\cite{MG-82}, the critical 
behavior of these materials should be described by the effective
Hamiltonian
\begin{equation}
{\cal H}_{H,rc} = - J \, \sum_{<xy>}  \rho_x \rho_y
\,\vec{s}_x \cdot \vec{s}_y  
+ a \sum_x \sum_i s_{x,i}^4 + \sum_x \sum_{ij} D_{x,ij} s_{x,i} s_{x,j},
\label{lattmodel-cubic-random}
\end{equation}
where, beside the random-exchange term, a random-anisotropy term is present.
Here the anisotropy term $D_{x,ij}$ is a random quantity that is traceless
and has zero average. Note that, for small anisotropy and weak disorder, 
this additional term should be smaller than the other ones, 
being, loosely speaking, proportional to the product of $(1-p)$ and $a$, i.e.,
it is a second-order perturbation of the Heisenberg Hamiltonian.
This argument is not fully justified at criticality, since, if  
random anisotropy is relevant, it will eventually change the 
critical behavior. Nonetheless, we expect---as we shall see, 
experiments confirm this assumption---a large preasymptotic region
in which such term can be neglected. For this reason in this paper 
we will not consider the random-anisotropy term and we shall discuss
the critical behavior of the model with Hamiltonian 
(\ref{lattmodel-cubic-random}) with $D_{x,ij} = 0$.

If only the random-exchange term is present, i.e. $D_{x,ij} = 0$, 
the critical behavior of the 
model (\ref{lattmodel-cubic-random}) can by studied using the
field-theoretical Hamiltonian
\begin{equation}
{\cal H}_{rc} = \int d^d x 
\left\{ {1\over 2}(\partial_\mu \phi(x))^2 + {1\over 2} r \phi(x)^2 + 
{1\over 2} \psi(x) \phi(x)^2 + 
{1\over 4!} v \left[ \phi(x)^2\right]^2 +
{1\over 4!} w \sum_{i=1}^M \phi_i(x)^4 \right\},
\label{Hphi4ranc}
\end{equation}
where $\psi(x)$ is a spatially uncorrelated 
random field with Gaussian distribution coupled to the 
local energy density.  Using the standard replica trick, one obtains the
Hamiltonian \cite{Aharony-75,tetra}
\begin{equation}
{\cal H}_e = \int d^d x 
\left\{ \sum_{i,a}{1\over 2} \left[ (\partial_\mu \phi_{a,i})^2 + 
         r \phi_{a,i}^2 \right] + 
\sum_{ij,ab} {1\over 4!}
\left( u + v \delta_{ij} +w \delta_{ij}\delta_{ab}\right)
          \phi^2_{a,i} \phi^2_{b,j} 
\right\},
\label{Hphi4}
\end{equation}
where $a,b=1,...M$ and $i,j=1,...N$.
The original system, i.e.   
the randomly dilute $M$-component cubic model, is recovered 
in the $N\rightarrow 0$ limit.\cite{assumption} 
The coupling $u$ is negative, being proportional to minus the variance of
the quenched disorder.

The study of the effective
Hamiltonian ${\cal H}_e$ in the limit $N\rightarrow 0$
provides also information on the critical behavior of
the randomly dilute antiferromagnetic $q$-state Potts model 
for $q=2$ and 3, with Hamiltonian
\begin{equation}
{\cal H}_{dq} = J \, \sum_{<xy>}  \rho_x \rho_y \delta_{s_x,s_y},
\label{dPotts}
\end{equation}
where $J>0$, $s_x=1,\ldots,q$, and
$\rho_x$ are uncorrelated random variables.
Indeed, as argued in Refs.~\cite{BGJ-80,Itakura-99} using RG arguments,
the critical behavior of the
antiferromagnetic three- and four-state 
Potts models on a cubic lattice
should be described by the cubic Hamiltonian ${\cal H}_c$ with
$M=2$ and $M=3$ respectively and with $w<0$.
The same correspondence holds in the random case. 
The randomly dilute three- and four-state models are respectively
related to the two- and three-component model with Hamiltonian 
(\ref{Hphi4}) in the limit $N\rightarrow 0$ and for $w<0$.

Since disorder is coupled to the local energy density,
one can use the Harris criterion 
\cite{Harris-74} to predict the critical behavior of the model.
It states that the addition of 
impurities to a system that undergoes a second-order 
phase transition does not change the critical behavior 
if the specific-heat critical exponent $\alpha_{\rm pure}$ of the pure 
system is negative. If $\alpha_{\rm pure}$ is positive, the transition
is altered. This occurs in the Ising case ($M=1$),
where the addition of impurities leads to a new random 
Ising universality class (RIM).
In pure $M$-component cubic models ($M>1$)
the specific-heat exponent $\alpha_{\rm pure}$ is negative; 
therefore, according to the Harris criterion, 
the pure fixed point is stable against disorder.
Nonetheless, disorder may still have physical consequences. 
For instance, it may give rise to new fixed points
or change the attraction domain of the pure stable 
fixed point. Systems that are outside the attraction domain of the fixed point
in the absence of disorder, and therefore
show a fluctuation-driven first-order transition,
may undergo a {\em second}-order transition in the presence of disorder.
Such phenomenon, usually referred to as softening, is well-understood in 
two-dimensional random-exchange models in which disorder is coupled to
the local energy density. Indeed, 
it was argued in Ref.~\cite{IW-79}, and later put on a rigorous basis
\cite{AW-89,HB-89}, that
in two dimensions thermal first-order transitions 
become continuous in the presence
of quenched disorder coupled to the local energy density.
For the cubic model (\ref{Hphi4cubic}) for $v<0$, it was shown \cite{Cardy-96}
that such a softening persists in $2 + \varepsilon$ dimensions, 
while it is absent near four dimensions,
see the $\epsilon$-expansion analysis of
Sec.~\ref{sec3}. Thus, it is interesting to address this issue 
in three dimensions, where the analysis of Ref.\cite{IW-79} shows that
the occurrence of softening may depend on nonuniversal  
features of the model.

In order to study the RG flow of the effective Hamiltonian ${\cal H}_e$, 
we consider the  fixed-dimension perturbative method in $d=3$
and compute the RG functions perturbatively to six loops.
The analysis of such series allows us to determine the RG flow.
We briefly anticipate the main results of our analysis.
The stability of the stable fixed points of the pure theory
predicted by the Harris criterion is confirmed.
The region $v<0$ for any $M$ and  the region $w<0$ for $M\geq 3$ is
outside the attraction domain of the stable fixed point for
all physical values $u<0$. Moreover, for any $M$
there exists a fixed point in the RIM universality class, 
which is weakly unstable (i.e. with a very small crossover exponent) and may
give rise to observable crossover effects in physical systems. We do not find
fixed points in the region  $v<0$ for any $M$ and 
in the region $w<0$ for $M\geq 3$. Therefore, no softening is
expected, at least for sufficiently low impurity concentration 
to justify the field-theoretical approach.  As for the three-state random
Potts model, it  is expected to have an XY transition as in the pure case
\cite{threestate} or a first-order transition, depending on the 
value of the effective negative coupling $w$. The four-state random
antiferromagnetic Potts model is expected to
undergo a weak first-order transition.

We predict that cubic magnets with small positive anisotropy 
have a critical behavior controlled by the pure cubic fixed point, 
which has critical exponents very close to the
Heisenberg ones. Therefore,
experiments should effectively observe the standard 
O(3) critical exponents. This is in good agreement with the experiments 
that observe in most of the cases O(3) behavior with good accuracy,
see, e.g., Refs.~\cite{PV-r,Kaul-85}. On the other hand, systems 
that tend to magnetize along the cubic axes, should show a first-order 
transition, as in the pure case. Experimentally, such a transition 
has never been observed, probably because of the smallness of the 
cubic anisotropy: due to the very small crossover exponent, $\phi\approx 0.01$,
see, e.g., Refs. \cite{CPV-00,PV-r},
the cubic breaking can be observed only 
very close to the critical point, i.e. for $|T-T_c|/T_c \ll 10^{-4}$, 
which is the limit of most of the experiments. 
As already observed in Ref. \cite{BK-97}, {\em a posteriori}, 
the experimental results also confirm the validy of neglecting the 
random-anisotropy term in Eq. (\ref{lattmodel-cubic-random}) 
in the experimentally relevant range of parameters. Indeed, its
presence would give rise to a crossover to a first-order transition 
in all cases \cite{MG-82}. 
The associated crossover exponent is 
$\phi = 2 \phi_H - 2 + \alpha_H \approx 0.39$ 
%% phi_H = 1.260(11) lavoro multicritico
%% alpha_H = -0.1336(15)
%% phi = 0.386(22)
where $\phi_H$ and $\alpha_H$ are the quadratic-anisotropy and 
specific-heat exponents for the Heisenberg model. Such an 
exponent is sizable and thus one should have been able to observe the 
random anisotropy if it were not very small.

The paper is organized as follows.
In Sec. \ref{sec2} we discuss some general properties of the RG flow
in three dimensions.
Sec. \ref{sec3} analyzes the RG flow  near four dimensions.
In Sec.~\ref{sec4} we present the computation and the analysis
of the fixed-dimension pertubative expansion to six loops.
In the Appendix we compute the differences between the three-component 
cubic and Heisenberg critical exponents, by a reanalysis of  the fixed-dimension
six-loop expansion of Ref. \cite{CPV-00} 
for the cubic Hamiltonian ${\cal H}_c$.

\section{General considerations on the RG flow}
\label{sec2}

In this section we discuss some properties
of the RG flow of the Hamiltonian (\ref{Hphi4}) for $N\rightarrow 0$,
using general arguments and known results holding
for the special cases in which one of the quartic couplings vanishes.

\begin{figure}[t]
\vspace{-2truecm}
\hspace{-1truecm}\centerline{\psfig{width=13truecm,angle=-90,file=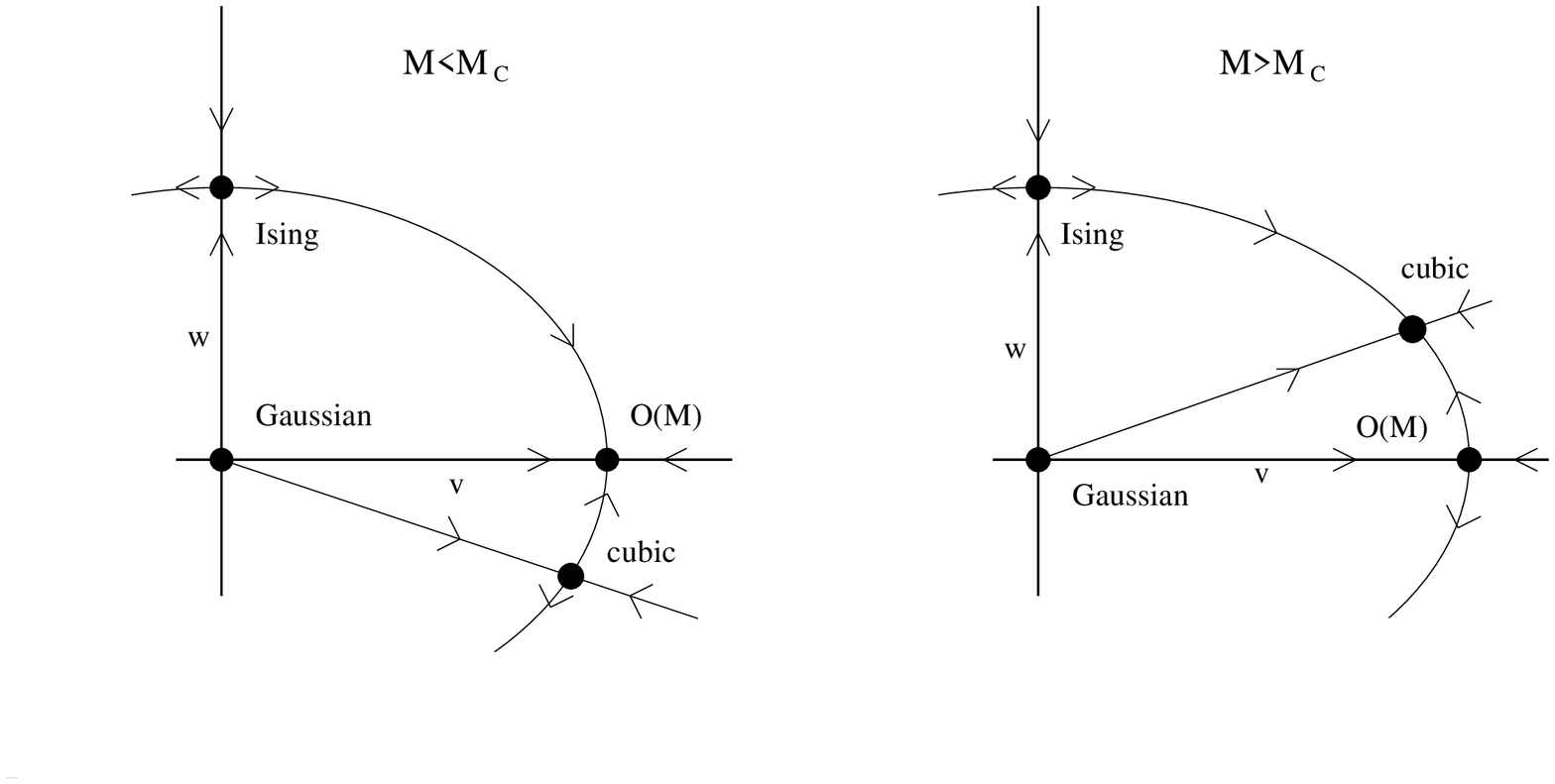}}
\vspace{-4truecm}
\caption{RG flow in the plane $u=0$ for $M<M_c$ and $M>M_c$.}
\label{flowu0}
\end{figure}

The RG flow in the plane $u=0$ is that of the
cubic Hamiltonian (\ref{Hphi4cubic}) (see, e.g., Ref.~\cite{PV-r}
for a recent review of results). Indeed,
for $u=0$ the Hamiltonian (\ref{Hphi4}) describes $N$ decoupled
$M$-component cubic-symmetric models. 
Therefore, in the plane $u=0$ there are four fixed points \cite{Aharony-76}:
the trivial Gaussian one, the Ising one in which the $M$ components of the 
field decouple,
the O($M$)-symmetric and the  cubic fixed point. 
The Gaussian fixed point is always unstable, 
and so is the Ising fixed point for any number of components $M$ 
\cite{Aharony-76,Sak-74}. 
On the other hand,
the stability properties of the O($M$)-symmetric and of the cubic
fixed point depend on $M$.
For sufficiently small values of $M$, $M<M_c$, the 
O($M$)-symmetric fixed point is stable and the cubic one is unstable.
For $M>M_c$, the opposite is true:
the RG flow is driven towards the cubic fixed point.
Figure~\ref{flowu0} sketches the 
flow diagram in the two cases $M<M_c$ and $M>M_c$.
High-order perturbative computations in the $\epsilon$-expansion and
in the fixed-dimension field-theoretical frameworks show that $2<M_c<3$;
more precisely, $M_c \approx 2.9$ \cite{FHY-00,CPV-00,Varnashev-00}.
This means that the critical behavior of the two-component cubic model is
described by the O(2)-symmetric fixed point and
therefore belongs to the XY universality class.
If $M>M_c$, the cubic anisotropy is relevant and therefore the critical
behavior of the system is not described by the Heisenberg isotropic 
Hamiltonian. 

In the three-component case
the cubic  critical exponents $\nu_c$, $\eta_c$ differ very little 
from those of the Heisenberg universality class.
Indeed, the analysis of the six-loop fixed-dimension expansions of 
Ref. \cite{CPV-00} reported in the Appendix 
provides the following estimates for their differences
\begin{equation}
\nu_{c} - \nu_H = -0.0003(3) ,\quad
\eta_{c} - \eta_H = -0.0001(1),\quad
\gamma_{c} - \gamma_H = -0.0005(7) . \label{diffexp}
\end{equation}
Note that these differences are much smaller than the 
typical experimental errors, see, e.g., Ref.~\cite{PV-r}
for a list of experimental results, so that, at present, cubic effects 
are experimentally negligible.
Using the accurate estimates of Ref.~\cite{CHPRV-02} for the Heisenberg
exponents and Eq.~(\ref{diffexp}), one obtains
\begin{equation}
\nu_c=0.7109(6), \qquad \eta_c=0.0374(5), \qquad \gamma_c=1.3955(12),
\label{cubicexp}
\end{equation}
which are consistent with, but much more precise than, the results obtained
from a direct analysis, see, e.g., Refs.~\cite{CPV-00,FHY-00}.

The stability of the pure fixed points against the $u$-perturbation
can be inferred by using general arguments \cite{Harris-74,Sak-74,Aharony-76}.
Since the $u$-interaction is the sum of the products 
of the energy operators of the different cubic $M$-component models,
the crossover exponent associated with $u$
is given by the specific-heat critical exponent $\alpha$ independently of $N$,
and thus also for $N\rightarrow 0$.
Therefore, the pure stable fixed point is 
stable with respect to random dilution for any $M\geq 2$, 
since the specific-heat exponent is always negative.
For example, for $M=2$, where the stable fixed point
is the O(2)-symmetric one, we have \cite{CHPRV-01}
$\alpha_{\rm XY}=-0.0146(8)$;
for $M=3$, where the stable fixed point is the cubic one,
$\alpha_c=-0.133(2)$ using Eq.~(\ref{cubicexp}).

\begin{figure}[t]
\hspace{-1truecm}\centerline{\psfig{width=7truecm,angle=-90,file=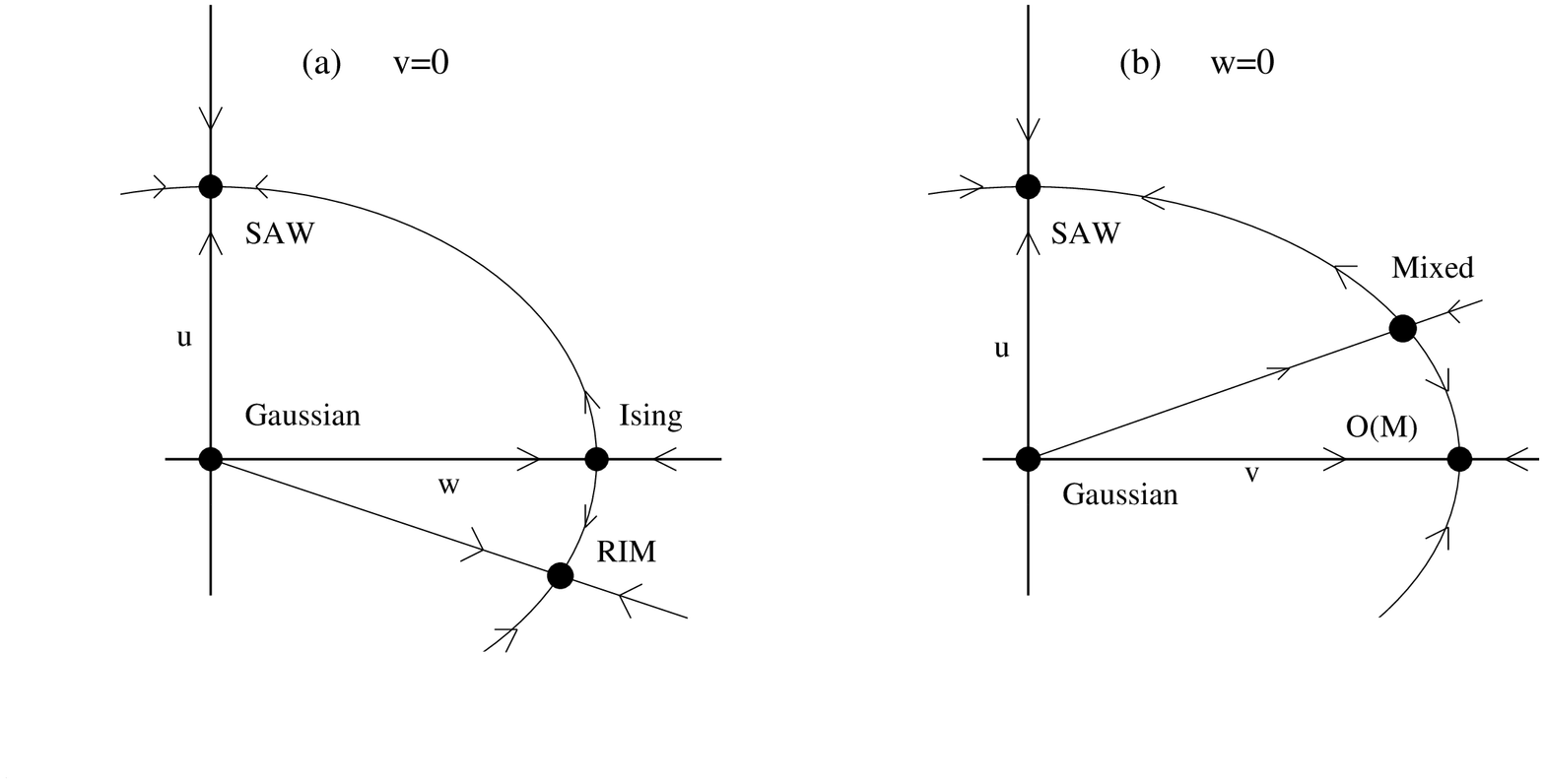}}
\vspace{-1truecm}
\caption{RG flow in the planes $v=0$ and $w=0$ for $N=0$.}
\label{flowvw0}
\end{figure}

For $v=0$ the Hamiltonian (\ref{Hphi4})
describes an $MN$-component model with cubic anisotropy.
The RG flow for $N\rightarrow 0$ is shown in Fig.~\ref{flowvw0}.
It is characterized by the presence of two stable fixed points. 
The one for $u>0$, $w=0$ is in the
self-avoiding walk 
(SAW) universality class, but it is irrelevant for our problem, since
it is unreachable from the physical region $u<0$.
The other one belongs to the region $u<0$, $w>0$ and it is in
the RIM universality class. See, e.g., 
Refs.~\cite{BFMMPR-98,FHY-00-2,PS-00,PV-00,TMVD-01}
for recent studies of the critical properties of RIM.

In the case $w=0$, the Hamiltonian (\ref{Hphi4}) describes $N$ coupled
$M$-vector models, and it is also called $MN$ model \cite{Aharony-76}.
See, e.g., Ref.~\cite{PV-r} for a recent review of results.
The RG flow for $M\geq 2$ and $N\rightarrow 0$ is shown in Fig.~\ref{flowvw0}.
Again, the flow is characterized by two stable fixed points:
the SAW and the O($M$)-symmetric ones.

For $M=2$ and generic $N$ the Hamiltonian (\ref{Hphi4})
is invariant under the transformation
\begin{eqnarray}
(\phi_{1,i}\;,\;\phi_{2,i}) &\longrightarrow&  {1\over \sqrt{2}}
(\phi_{1,i}+\phi_{2,i}\;,\;\phi_{1,i}-\phi_{2,i}),
\nonumber \\
(u_0,v_0,w_0) &\longrightarrow& (u_0,v_0+\case{3}{2}w_0,-w_0).
\label{sym1}
\end{eqnarray}                                              
For $N=0$ this transformation maps the Ising fixed point into the cubic one,
and  the RIM fixed point into a new
one belonging to the region with $u<0$, $v>0$, $w<0$.
Of course, corresponding fixed points describe  the same critical behavior.

In conclusion, the above-reported considerations show the presence of
at least seven fixed points for $M\geq 3$ 
and eight for $M=2$ using the above-mentioned symmetry.
Of course, other fixed points may lie outside the planes
$u=0$, $v=0$, $w=0$. Thus, a more general analysis for
generic values of the quartic couplings
is necessary in order to obtain a satisfactory knowledge of the RG flow.
In particular, this would allow us to investigate
if random dilution may cause a softening 
of the first-order transition predicted for pure systems that are
outside the attraction domain of the stable fixed point.
For example, systems with $v<0$ might have 
a continuous transition if there exists a fixed point 
in the region $u<0$, $v<0$, while, for systems with $M\geq 3$ and $w<0$,
softening to a continuous transition
requires a fixed point in the region $u<0$, $w<0$.

\section{The RG flow near four dimensions}
\label{sec3}

The RG flow of the randomly dilute cubic-symmetric model can be investigated
near four dimensions using the first nontrivial terms of the
expansion in powers of $\epsilon\equiv 4-d$.
Using the results reported
in Refs.~\cite{Aharony-75,JK-77,SAS-97},
one may easily obtain the results of Table~\ref{epsexp},
where the location of the fixed points 
and the eigenvalues of the corresponding stability matrix
are reported to leading order.

\begin{table}[tbp]
%\squeezetable
\caption{
Fixed points of the Hamiltonian (\ref{Hphi4}) near four dimensions.
We report the leading nontrivial contribution of the expansion in powers
of $\epsilon$. Note that $K_d = (4\pi)^d \Gamma(d/2)/2$.
}
\label{epsexp}
\begin{tabular}{cccccl}
\multicolumn{1}{c}{}&
\multicolumn{1}{c}{}&
\multicolumn{1}{c}{$u/K_d$}&
\multicolumn{1}{c}{$v/K_d$}&
\multicolumn{1}{c}{$w/K_d$}&
\multicolumn{1}{c}{stability eigenvalues}\\
\tableline \hline
I  & Gaussian      & 0 & 0                       & 0
& $\omega_u = -\epsilon$, $\omega_v = -\epsilon$, $\omega_w = -\epsilon$ \\

II & O($M$)        & 0 & ${6\over M+8} \epsilon$ & 0 
& $\omega_u = {4-M\over M+8}\epsilon$, 
$\omega_v = \epsilon$, $\omega_w = {4-M\over M+8}\epsilon$ \\

III& SAW           & ${3\over 4}\epsilon$ &0&0
& $\omega_u = \epsilon$, 
$\omega_v = \epsilon/2$, $\omega_w = \epsilon/2$ \\

IV & Ising         & 0 & 0 & ${2\over 3}\epsilon$
& $\omega_u = -\epsilon/3$, 
$\omega_v = \epsilon$, $\omega_w = - \epsilon/3$ \\

V  & Mixed         
& ${M-4 \over M-1}{3\over 8}\epsilon$ & ${3\over 2(M-1)}\epsilon$ & 0
& $\omega_1 = \epsilon$, 
$\omega_2 = {4-M\over 3(M+1)}\epsilon$,	
$\omega_w = {4-M\over 3(M+1)}\epsilon$ \\

VI  & Cubic         & 0& ${2\over M} \epsilon$& ${2(M-4)\over 3M}\epsilon$
& $\omega_u = {M-4\over 3 M}\epsilon$, 
$\omega_2 = \epsilon$, $\omega_3 = {M-4\over 3M}\epsilon$ \\

VII & RIM          & $-2\sqrt{27 \over 106} \sqrt{\epsilon}$ & 0&
$2\sqrt{24 \over 53} \sqrt{\epsilon}$
& $\omega_1 = 2 \epsilon$, 
$\omega_v = -\sqrt{24\over 53}\sqrt{\epsilon}$, 
$\omega_3 = \sqrt{24\over 53}\sqrt{\epsilon}$\\

VIII &          & $2\sqrt{27\over 106} \sqrt{\epsilon}$ & 0&
$-2\sqrt{24\over 53} \sqrt{\epsilon}$ 
& $\omega_1 = O(\epsilon)$, 
$\omega_v = -\sqrt{24\over 53}\sqrt{\epsilon}$, 
$\omega_3 = \sqrt{24\over 53}\sqrt{\epsilon}$\\

IX & ($M>2$)  & ${M-4 \over 4(M-2)}\epsilon$ &
${1\over(M-2)}\epsilon$ & ${M-4 \over 3(M-2)}\epsilon$ 
& $\omega_1 = \epsilon$, 
$\omega_2 = {4-M\over 6(M-2)}\epsilon$, $\omega_3 = {M-4\over 6(M-2)}\epsilon$ \\

IX, X & ($M=2$)  & $\mp 2\sqrt{27\over 106} \sqrt{\epsilon}$ & 
$\pm 2\sqrt{54\over 53} \sqrt{\epsilon}$& $\mp 2\sqrt{24\over 53}
\sqrt{\epsilon}$ & \\
\end{tabular}
\end{table}

Note that the 
$O(\sqrt{\epsilon})$ fixed points emerge at two-loop level,
and are related to the degeneracy  of the one-loop $\beta$-functions \cite{Kh-75}.
For $M=2$ the fixed points IV, VII, VIII can be
mapped  respectively 
into the fixed points VI, IX, X  using the symmetry (\ref{sym1}).

The physically relevant fixed points are those 
that can be reached from the region $u<0$, i.e. the fixed points
I (Gaussian), II (O($M$)), IV (Ising), VI (cubic), 
VII (RIM) for all values of $M\geq 2$, and 
V (mixed) and  IX for $M\leq 4$
(note that in three dimensions the mixed fixed point V
lies in the $u>0$ region for all $M\geq 2$, as shown in 
the preceding section).
Concerning their stability properties,
we note that all fixed points in the region $u\leq 0$, except 
the $O(M)$ and cubic one, 
are unstable for any $M$.
For $M<M_c=4-2\epsilon+O(\epsilon^2)$ ($M_c\approx 2.9$ in three dimensions)
the stable fixed point is the O($M$)
one, while for $M>M_c$ the stable fixed point is the cubic one.
Note that the derivation of $\omega_1$ for the fixed point
VII (RIM) requires a three-loop calculation---we used here
the results of Ref.~\cite{JK-77}.

In conclusion, near four dimensions 
the critical behavior is not changed by 
the addition of random impurities for any $M\geq 2$.
Moreover, as already observed in Ref.~\cite{Cardy-96}
(although that analysis missed the $O(\sqrt{\epsilon})$
fixed points),
there is no  softening of the transition for pure systems
that are outside the attraction domain
of the stable fixed point (in particular, for $v<0$ and 
for $w<0$ in the case $M\geq M_c$).

Even though the $\epsilon$ expansion provides useful indications of the 
RG flow in lower dimensions, the validity of the extrapolation to $\epsilon=1$
of the results obtained near four dimensions is not guaranteed,
even at a qualitative level.
Relevant features concerning the location and the stability of
the fixed points can drastically change approaching three dimensions.
Moreover,  new fixed points, 
which are not found in $\epsilon$-expansion analyses, may appear in 
three dimensions.
For example, this occurs in
the physically interesting cases 
of the $\varphi^4$ Hamiltonian describing
the critical behavior of frustrated spin models with noncollinear
order, see. e.g., Refs.~\cite{Kawamura-98,PV-r}, and the Ginzburg-Landau
model of superconductors,  where
a complex scalar field couples to a gauge field,
see, e.g., Ref.~\cite{superc}.

\section{Analysis of the six-loop fixed-dimension perturbative expansion.}
\label{sec4}

\subsection{The six-loop series}
\label{sec41}

In the fixed-dimension field-theoretical approach, see, e.g., Ref.~\cite{PV-r}
and references therein,
one expands in powers of appropriately defined zero-momentum
quartic couplings.  
In the present case we define renormalized couplings 
$\ub$, $\vb$, and $\wb$ from the zero-momentum four-point
function. They are normalized so that, at tree level, they are related to the 
bare couplings $u$, $v$, and $w$ by
\begin{equation}
u = {16\pi\over3} R_{MN} \ub m, \quad
v = {16\pi\over3} R_{M}  \vb m, \quad
w = {16\pi\over3}   \wb m, 
\end{equation}
where $R_K = 9/(8+K)$ and  $m$ is the renormalized mass defined by
\begin{equation}
\Gamma^{(2)}_{ai,bj}(p) =
  \delta_{ai,bj} Z_\phi^{-1} \left[ m^2+p^2+O(p^4)\right].
\label{ren1}  
\end{equation}
Here $\Gamma^{(2)}_{ai,bj}(p)$ is
the two-point one-particle irreducible correlation function.

By using the numerical results compiled in Ref.~\cite{NMB-77} and 
a general symbolic manipulation program, we determined the RG functions
to six loops.
The resulting series for $N=0$ are
\begin{eqnarray}
&&\beta_u  =
m{\partial \ub\over \partial m}=
-\ub + \ub^2 + {2(M+2) \over (M+8)} \ub \, \vb + {2\over 3} \ub \, \wb 
- {95\over 216} \ub^3 - {50 (M+2) \over 27(M+8) } \ub^2 \, \vb
- {92(M+2) \over 27(M+8)^2} \ub\, \vb^2 \nonumber \\ 
&&-{50 \over 81 }\ub^2 \, \wb 
- {92 \over 729 } \ub\, \wb^2 
- {184 \over 81(M+8)} \ub\,\vb\,\wb 
+\ub \;\left(\sum_{i+j+k\geq 3} b^{(u)}_{ijk} \;\ub^i\, \vb^j \,\wb^k \right),
\label{bu}
\end{eqnarray}
\begin{eqnarray}
&&\beta_v  =  m{\partial \vb\over \partial m}=
-\vb + \vb^2 + 
{3\over 2} \ub \, \vb + {2\over 3}\vb\, \wb 
- {4(190+41 \,M) \over 27 (M+8)^2} \vb^3 
- {2(131+25 \,M) \over 27  (M+8)} \ub\, \vb^2 \nonumber \\ 
&&- {185 \over 216} \ub^2 \, \vb 
- {400 \over 81 (M+8)} \vb^2 \, \wb 
- {92 \over 729} \vb \,  \wb^2  
- {77\over 81} \ub\,\vb\,\wb 
+\vb \;\left(\sum_{i+j+k\geq 3} b^{(v)}_{ijk} \;\ub^i \, \vb^j \, \wb^k \right),
\label{bv}
\end{eqnarray}
\begin{eqnarray}
&&\beta_w  =  m{\partial \vb\over \partial m}=  -\wb + \wb^2  
+ {3\over 2 }\ub\, \wb + {12\over (M+8)}\vb\, \wb  
- {308 \over 729} \wb^3 - {104 \over 81} \ub\, \wb^2  
- {832 \over 81(M+8)} \vb\, \wb^2 
\nonumber \\ 
&&- {555 \over 648} \ub^2 \, \wb
- {4(370+23 M) \over 27(M+8)^2} \vb^2 \, \wb 
- {(23 M+370)\over 27(M+8)} \ub\,\vb\,\wb  
+\wb \;\left(\sum_{i+j+k\geq 3} b^{(w)}_{ijk} \;\ub^i \, \vb^j \, \wb^k \right).
\label{bw}
\end{eqnarray}
The coefficients 
$b^{(u)}_{ijk}$, $b^{(v)}_{ijk}$, $b^{(w)}_{ijk}$, with $3\leq i+j+k\leq  6$, 
are reported in the Tables~\ref{betauc}, \ref{betavc}, and \ref{betawc},
respectively.
We have also computed the RG functions associated with the
critical exponents
to six loops. We do not report them either since they will not be used in our
analysis  (they are available on request).

\subsection{Resummation of the series}
\label{sec42}

The perturbative series for the $M$-vector model and for the cubic model are 
Borel summable and thus accurate results can be obtained by resummation
methods that exploit
Borel summability
and the knowledge of the large-order behavior \cite{ZJ-book,CPV-00}.
On the other hand, perturbative series for randomly dilute models cannot 
be resummed naively. In the present case, an extension of the results 
of Refs. \cite{BMMRY-87,McKane-94} indicates that the perturbative series 
at fixed $u/v$ and $u/w$ are not Borel summable. 
For the zero-dimensional random Ising model, Ref. \cite{AMR-99} showed
that one could still compute correctly the free energy 
by means of a more elaborate resummation method.
Although no proof exists
that this procedure works in higher dimensions,
this method was recently applied to resum  six-loop
expansions for the random Ising model in three dimensions \cite{PV-00},
obtaining reasonably accurate results.

Here,
we apply a similar method to resum the perturbative series for the
randomly dilute cubic model for generic values of $u,v,w$.
If $f(u,v,w)$ has a perturbative expansion of the form
\begin{eqnarray}
&&f(u,v,w)= \sum_{n=0} c_n(v,w) u^n,\label{f1}\\
&&c_n(v,w)= \sum_{k,l=0}  c_{nkl} v^k w^l \label{f2},
\end{eqnarray}
following Ref.~\cite{PV-00}, we first resum
the coefficients $c_n(v,w)$ and then, using the computed coefficients, we resum
the series in $u$. 
The first resummation can be performed by using either the 
Pad\'e-Borel method or the conformal-mapping method---indeed, the 
large-order behavior of the coefficients $c_n(v,w)$ is exactly the one of the 
$M$-component cubic model~\cite{CPV-00}.
In the Pad\'e-Borel method, for each $0\le n\le p$ (where $p$ is
maximum order considered), we consider 
the Pad\'e approximants $[(p-n-r_n)/r_n]$ of the Borel-Leroy transformed
series, which depend on an additional parameter $b_n$.
The second method uses the large-order behavior of the series 
and a conformal mapping, see Ref. \cite{PV-00}. 
Once an estimate of the coefficients $c_n(v,w)$ is obtained, 
the second resummation is performed by using the  Pad\'e-Borel method.
We consider Pad\'e approximants $[(q-r_u)/r_u]$ of the Borel-Leroy 
transformed series, with $q\leq p-1$.
Therefore, in the double Pad\'e-Borel method, which we will mostly use,
the parameters introduced by the analysis are 
$p,\,\{b_n\},\,\{r_n\}$ for the 
first resummation and $q,\,b_u,\,r_u$ for the second one.
It is impossible to vary all the $b_n$ and $r_n$ independently, since there are
too many combinations. Therefore, following Ref.~\cite{PV-00},
we took them equal for all $n$.

In the special cases when one of the coupling vanishes,
one may use the methods already applied in the literature 
for the study of the corresponding model.
For example, the series for $u=0$ correspond to those of
the Hamiltonian (\ref{Hphi4cubic}); they are
Borel summable, 
so that one may use the standard technique based on 
the knowledge of the large-order behavior and a
conformal mapping \cite{CPV-00}.
The series for $v=0$ correspond to those of the
random Ising Hamiltonian (\ref{Hphi4cubic}) with $M\rightarrow 0$;
thus, one may use the analysis methods outlined in Ref.~\cite{PV-00}.

\subsection{Results}
\label{sec43}

\subsubsection{Stability of the fixed points of the pure systems}
\label{sec431}

First, we check the stability of the stable fixed point of the pure
theory $(\overline{u} = 0$) 
with respect to the perturbation induced by random dilution.

The coordinates of the stable O(2)-symmetric fixed point
of the two-component theory are  \cite{CHPRV-01,GZ-98}
$\overline{u}=0$, $\overline{v}_{XY}=1.402(4)$,
and $\overline{w}=0$.
The stability with respect to random dilution is controlled by 
\begin{equation}
\omega_u =  \left. 
{\partial \beta_u \over \partial \ub}\right|_{(0, \vb_{XY},0)},
\end{equation}
which, according to the nonpertubative 
argument reported in Sec.~\ref{sec2}, should be given by 
\begin{equation}
\omega_u= -{\alpha_{XY}\over \nu_{XY}},
\label{omegau}
\end{equation} 
where $\alpha_{XY}$ and $\nu_{XY}$ are the
critical exponents of the $XY$ model. 
The analysis of the series (exploiting the known large-order
behavior of the series and using the conformal-mapping method) gives 
$\omega_u = 0.007(8)$.
Therefore, the stability of the $XY$ fixed point is substantially confirmed,
although the apparent error of the analysis does not completely
exclude the opposite sign for $\omega_u$.
The estimate of $\omega_u$
is substantially consistent with Eq.~(\ref{omegau}).
Indeed, $\alpha_{XY}/\nu_{XY} = - 0.0217(12)$ (Ref.~\cite{CHPRV-01}), 
obtained from the analysis of high-temperature series,
and \cite{GZ-98} $\alpha_{XY}/\nu_{XY} = -0.016(7)$,
obtained by a more similar technique, i.e., the
analysis of the fixed-dimension expansion of the O(2)-symmetric model.
We mention that perturbative studies based on shorter
series \cite{MV-01,SV-99}
give apparently contradictory results, favoring
a negative value of $\omega_u$.

We now consider the case $M\geq 3$. The relevant
fixed point is now the cubic one with coordinates $(0,\vb_c,\wb_c)$.
The stability against the $\ub$ perturbation is determined by the sign of
\begin{equation}
\omega_u= \left. {\partial \beta_u \over \partial
\ub} \right|_{(0,\vb_c,\wb_c)}.
\end{equation}
Again, one expects 
$\omega_u = -{\alpha_c/\nu_c}$,
where $\alpha_c$ and
$\nu_c$ are the critical exponents associated with the
cubic fixed point.
Since the large-order behavior of the expansion of $\omega_u$ is that 
of the series of the cubic-symmetric pure model,
one can use the standard conformal-mapping resummation~\cite{CPV-00}. 
Estimates of $\omega_u$ for several values of $M$  are reported 
in Table \ref{omu}~(we use the estimates of the 
cubic fixed points reported in  Ref. \cite{CPV-00}).
For comparison, we also report the ratio $-\alpha_c/\nu_c$ as obtained 
from the results of Ref. \cite{CPV-00}.
Other estimates can be obtained using the results of Refs.~\cite{FHY-00,MV-98}.
For $M=3$, we also quote $-\alpha_c/\nu_c \approx 0.142$ \cite{MV-98},
$-\alpha_c/\nu_c \approx 0.163$ \cite{FHY-00}.

\begin{table}[tbp]
\caption{
Estimates of the subleading exponent $\omega_u$ at the $M$-component 
cubic fixed point for several values of $M$. 
The last column reports the theoretical prediction $-\alpha_c/ \nu_c$
obtained from the results of Ref.~\protect\cite{CPV-00}.
For $M=3$ we also report the number obtained from the estimates
(\protect\ref{cubicexp}).
}
\label{omu}
\begin{tabular}{clll}
\multicolumn{1}{c}{$M$}&
\multicolumn{1}{c}{$[\vb_c,\wb_c]$}&
\multicolumn{1}{c}{$\omega_u$}&
\multicolumn{1}{c}{$-\alpha_c/\nu_c$}\\
\tableline \hline
$3$      & $[1.321(18),0.096(20)]$ & $0.157(28)$ & $0.167(24)$,$\;\;0.187(3)$ \\
$4$      & $[0.881(14),0.639(14)]$ & $0.203(27)$ & $0.199(30)$ \\
$8$      & $[0.440(12),1.136(10)]$ & $0.199(21)$ & $0.191(24)$ \\
$\infty$ & $[0.174(6) ,1.417(6) ]$ & $0.179(16)$ & $0.175(30)$ \\
\end{tabular}
\end{table}

In conclusion, the analysis of the six-loop perturbative 
series shows that the stable fixed point of the system without disorder
is stable with respect to the addition of random impurities,
in agreement with the Harris criterion.

\subsubsection{Fixed points induced by the disorder, i.e. for $u<0$.} 
\label{sec432}

We now study  the stability of the RIM fixed point 
located in the plane $v=0$. Its coordinates are \cite{PV-00}
$\ub_{\rm RIM}=-0.631(16)$, $\vb = 0$, $\wb_{\rm RIM}=2.195(20)$.
This fixed point is stable in the plane $v=0$.
To check its stability with respect to the $v$-perturbation 
we need to compute only
\begin{equation}
\omega_v=\left.{\partial \beta_v \over \partial \vb}
\right|_{(\ub_{\rm RIM},0,\wb_{\rm RIM})}\, ,
\end{equation}
since the derivatives of $\beta_v$ with respect to $\ub$ and $\wb$ vanish
at $\vb=0$. 
Note that the series for $\omega_v$ is $M$ independent.
We apply the double Pad\'e-Borel method to resum the 
series. We find that nondefective Pad\'e 
approximants are obtained only for 
$r_w=1$. 
Setting $q=p-1$, the results depend on 
four free parameters: $p$, $r_u$, $b_u$, and $b_w$. In principle,
we should look for the values of $b_u$ and $b_w$ that make the estimates
independent of the order $p$ of the series, but in the present case the 
results are weakly depending upon these two parameters. 
So, we consider all values $0\leq b_u, \, b_w \leq 10$.
With this choice, all approximants 
with $r_u=1$ and $r_u > 2$ are defective and therefore,
we use $p=4,\,5,$ and 6 and $r_u=0,\,2$.
For $r_u=0$~(direct summation) we obtain
$\omega_v = -0.047(8)\{20\}$ for $p=4$, 
$\omega_v=0.008(2)\{20\}$ for $p=5$,  and
$\omega_v=-0.049(2)\{20\}$ for $p=6$,
where the number between parentheses 
is related to the spread of the results
of the approximants considered, while the one between braces is related to
the uncertainty on the location of the fixed point.
For $r_u=2$, all approximants with $p=4$ are defective. 
For $p=5$ we obtain
$\omega_v=-0.012(14)\{20\}$ and for $p=6$ $\omega_v=-0.080(5)\{20\}$.
The quite large discrepancy between these estimates clearly 
indicates that the analysis is not very robust.
We give as conservative final estimate
$\omega_v=-0.04(5)$,
that includes all previous results~\cite{footnote4}.
This value suggests that the RIM fixed point is unstable,
although it does not allow us to exclude the opposite case.

Finally, we search for the presence of new fixed points in the physical 
region $\ub<0$.  
Our analysis does not provide evidence for new fixed points in the
physical region $\ub<0$, at least for $\ub \gtrsim -1 $, the region in which
sufficiently stable results are obtained.
In the case $M=2$, we only find the fixed point
predicted by the symmetry (\ref{sym1}). Indeed, a fixed point 
equivalent to the RIM is expected at
$\ub^*=\ub^*_{\rm RIM}=-0.631(16)$, 
$\vb^*=\vb^*_{\rm RIM}+{5\over 3} \wb^*_{\rm RIM} =3.658(33)$, 
$\wb^*=-\wb^*_{\rm RIM}=-2.195(20)$.
Our resummation of the six-loop series gives consistent 
results for the location of the fixed point, i.e.  
$\ub^*=-0.70(5)$, $\vb^*=3.7(3)$, $\wb^*=-2.2(2)$.

\section*{Acknowledgment}

We thank Yu. Holovatch for useful correspondence.

\appendix

\section{Differences between cubic and 
Heisenberg critical exponents}
\label{diffcH}

In the three-component case
the cubic  critical exponents differ very little 
from those of the Heisenberg universality class.
The available estimates of the critical exponents are
indistinguishable within their uncertainty.
Field-theoretical six-loop calculations for the cubic model
\cite{CPV-00} give $\nu_c=0.706(6)$, $\eta_c=0.0333(26)$ and
$\gamma_c=1.390(12)$, while the analysis of Ref.~\cite{GZ-98}
of six- and seven-loop series for the O(3)-symmetric Heisenberg model 
provide the estimates $\nu_H=0.7073(35)$, $\eta_H=0.0355(25)$,
$\gamma_H=1.3895(50)$. We also mention the more accurate estimates 
$\nu_H=0.7112(5)$, $\eta_H=0.0375(5)$, $\gamma_H=1.3960(9)$,
obtained by lattice techniques \cite{CHPRV-02}.
By comparing these estimates we can only put a bound
on the differences of the cubic and Heisenberg exponents; for example,
the estimates of $\nu$ differ at most by 1\%.

Much better estimates of such differences can be obtained 
by a more careful analysis of the six-loop fixed-dimension series
of the cubic model (\ref{Hphi4cubic}) computed in Ref.~\cite{CPV-00}. 
We recall that the expansion 
is performed in powers of the zero-momentum quartic couplings
$\vb$ and $\wb$ associated respectively with 
the couplings $v$, $w$ of the cubic Hamiltonian
(\ref{Hphi4cubic}).
The series for the cubic model
can be  obtained by setting $\ub=0$ in the series reported
in Sec.~\ref{sec4}.
In the plane $\vb,\wb$ the coordinates 
of the O(3)-symmetric and cubic fixed points
are respectively \cite{GZ-98,CHPRV-02}
$X_H=[1.390(4),0]$ and \cite{CPV-00}
$X_c=[1.321(18),0.096(20)]$.
In order to obtain a more precise determination of the
exponents, one may proceed as follows. 
Noting the closeness of the fixed points, one may
estimate the difference of the corresponding critical esponents
$\eta$ and $\nu$ by expanding around the
Heisenberg fixed point. Therefore, the first-order
approximation is
\begin{eqnarray} 
&&\Delta\eta \equiv \eta_c - \eta_H \approx \Delta\eta^{(1)}=
\left. {\partial \eta_\phi \over  \partial \vb} \right|_{X_{H}}
\left( \vb_c-\vb_H \right) +  
\left. {\partial \eta_\phi \over  \partial \wb} \right|_{X_H}
\wb_c, \label{linapprox} \\
&&\Delta\nu \equiv \nu_{c} - \nu_{H} \approx \Delta\nu^{(1)}=
\left. {\partial \nu\over  \partial \vb} \right|_{X_{H}}
\left( \vb_c-\vb_{H} \right) +   
\left. {\partial \nu\over  \partial \wb} \right|_{X_{H}} \wb_c. 
\nonumber
\end{eqnarray}
The RG functions $\eta_\phi(\vb,\wb)$, $\eta_t(\vb,\wb)$, and 
$\nu(\vb,\wb) = \left( 2 - \eta_\phi +
\eta_t\right)^{-1}$ have been introduced
in Ref.~\cite{CPV-00}.
The expressions (\ref{linapprox}) can be simplified using
the relation \cite{extc}
\begin{equation}
\left. {\partial \eta_{\phi,t}\over \partial \wb} \right|_{(\vb,0)}=
{M+8\over 3 (M+2)}
\left. {\partial \eta_{\phi,t}\over \partial \vb} \right|_{(\vb,0)},
\label{euevrel} 
\end{equation}
which is verified by our six-loop series and is probably exact.
Therefore, we write
\begin{eqnarray}
&&\Delta\eta^{(1)} =
\left. {\partial \eta_\phi \over  \partial \vb} \right|_{X_{H}}
\left( \vb_c-\vb_{H} + {11\over 15} \wb_c \right),   
\nonumber \\
&&\Delta\nu^{(1)} =
\left. {\partial \nu\over  \partial \vb} \right|_{X_{H}}
\left( \vb_c-\vb_{H} + {11\over 15} \wb_c \right).   \label{linapprox2}
\end{eqnarray}
In order to  
estimate the right-hand side of the above equations, 
one needs to consider all the elements of the
covariance matrix associated with
the cubic fixed point $X_c$, i.e.
$C_{\ub\,\ub} =0.000345$, $C_{\vb\,\vb}=0.000380$,
and $C_{\ub\,\vb} = -0.000361$,
because the estimates of the coordinates $\vb_c,\wb_c$ are strongly
correlated. 
Using the covariance matrix one obtains the quite
precise estimate
$\vb_c + {11\over 15} \wb_c  = 1.391(4)$,
which is very close to $\vb_{H}=1.390(4)$, 
thus leading to  a large cancellation in the right-hand side
of Eqs.~(\ref{linapprox2}).
In alternative, one may consider the approximate
relation
\begin{equation}
\left. {\partial \beta_{v}\over  \partial \vb} \right|_{X_{H}}
(\vb_c - \vb_{H}) + 
\left. {\partial \beta_{v}\over  \partial \wb} \right|_{X_{H}}
\wb_c \approx 0, 
\end{equation}
which is obtained by expanding the equation $\beta_v(\vb_c,\wb_c)=0$.
Within this approximation, one obtains
$\vb_c - \vb_{H} = -0.743(10) \;\vb_c$,
and therefore
$\vb_c-\vb_{H} + {11\over 15} \wb_c = -0.0009(8)$.
Moreover, the analysis of the series provides the results
$\left. {\partial \eta_\phi /  \partial \vb} \right|_{X_{H}}
= 0.06(1)$ and
$\left. {\partial \nu /  \partial \vb} \right|_{X_{H}} =0.21(1)$.
%\left. {\partial \eta_t \over  \partial \vb} \right|_{X_{H}} =-0.365(10)
Inserting in Eqs. (\ref{linapprox2}), we finally obtain
$\Delta \eta^{(1)} = -0.00005(5)$ and
$\Delta\nu^{(1)} = -0.0002(2)$.
Next, we determined the 
second-order contributions to Eq.~(\ref{linapprox}).
It can be estimated by
evaluating the second derivatives of $\eta_\phi$, $\eta_t$, and
$\beta_u$ at $X_H$. 
We obtained
$\Delta \eta^{(2)} \approx -0.00002$ and
$\Delta \nu^{(2)}\approx -0.0001$.
We also checked that the third-order contributions are 
very small and negligible.
Summing up, we obtain the estimates (\ref{diffexp}).

\begin{table}[tbp]
\vspace{0truecm}
\renewcommand\arraystretch{0.7}
\squeezetable
\caption{
The coefficients $b^{(u)}_{ijk}$, cf. Eq. (\ref{bu}).
}
\label{betauc}
\begin{tabular}{cl}
\multicolumn{1}{c}{$i,j,k$}&
\multicolumn{1}{c}{$R_M^{-j} b^{(u)}_{ijk}$}\\
\tableline \hline
0,0,3 &$0.090449$\\
0,1,2 &$0.266228 + 0.00511843\,M$\\
0,2,1 &$0.23096 + 0.0403865\,M$\\
0,3,0 &$0.0513245 + 0.034637\,M + 0.00448739\,M^2$\\
1,0,2 &$0.467389$\\
1,1,1 &$0.844729 + 0.0900481\,M$\\
1,2,0 &$0.281576 + 0.170804\,M + 0.015008\,M^2$\\
2,0,1 &$0.857364$\\
2,1,0 &$0.571576 + 0.285788\,M$\\
3,0,0 &$0.389923$\\
\hline
0,0,4 &$-0.0754467$\\
0,1,3 &$-0.298245 - 0.00354189\,M$\\
0,2,2 &$-0.434652 - 0.0180279\,M$\\
0,3,1 &$-0.24561 - 0.0570548\,M + 0.000878001\,M^2$\\
0,4,0 &$-0.040935 - 0.0299766\,M - 0.00460823\,M^2 + 0.0000731668\,M^3$\\
1,0,3 &$-0.476234$\\
1,1,2 &$-1.38978 - 0.0389194\,M$\\
1,2,1 &$-1.18263 - 0.250819\,M + 0.00474531\,M^2$\\
1,3,0 &$-0.262806 - 0.187141\,M - 0.0268143\,M^2 + 0.000527256\,M^3$\\
2,0,2 &$-1.2213$\\
2,1,1 &$-2.09844 - 0.344166\,M$\\
2,2,0 &$-0.699479 - 0.464461\,M - 0.0573609\,M^2$\\
3,0,1 &$-1.34386$\\
3,1,0 &$-0.895907 - 0.447954\,M$\\
4,0,0 &$-0.447316$\\
\hline
0,0,5 &$0.0874933$\\
0,1,4 &$0.434773 + 0.0026939\,M$\\
0,2,3 &$0.856086 + 0.0188466\,M$\\
0,3,2 &$0.805412 + 0.0689363\,M + 0.000584428\,M^2$\\
0,4,1 &$0.334915 + 0.0984617\,M + 0.00404184\,M^2 + 0.0000476511\,M^3$\\
0,5,0 &$0.0446554 + 0.0354559\,M + 0.00710302\,M^2 + 0.000275809\,M^3 + 
     3.17674\times {10}^{-6}\,M^4$\\
1,0,4 &$0.630396$\\
1,1,3 &$2.48473 + 0.0368538\,M$\\
1,2,2 &$3.52129 + 0.258766\,M + 0.00231537\,M^2$\\
1,3,1 &$1.95691 + 0.546998\,M + 0.0172536\,M^2 + 0.00042021\,M^3$\\
1,4,0 &$0.326152 + 0.254242\,M + 0.0484588\,M^2 + 0.00150784\,M^3 + 
     0.0000350175\,M^4$\\
2,0,3 &$1.93839$\\
2,1,2 &$5.51853 + 0.296649\,M$\\
2,2,1 &$4.61152 + 1.18578\,M + 0.0178836\,M^2$\\
2,3,0 &$1.02478 + 0.775897\,M + 0.135727\,M^2 + 0.00198707\,M^3$\\
3,0,2 &$3.14287$\\
3,1,1 &$5.26405 + 1.02168\,M$\\
3,2,0 &$1.75468 + 1.2179\,M + 0.17028\,M^2$\\
4,0,1 &$2.43619$\\
4,1,0 &$1.62413 + 0.812063\,M$\\
5,0,0 &$0.633855$\\
\hline
0,0,6 &$-0.117951$\\
0,1,5 &$-0.703992 - 0.00371266\,M$\\
0,2,4 &$-1.74808 - 0.0211855\,M $\\
0,3,3 &$-2.2733 - 0.0857507\,M + 0.0000391428\,M^2$\\
0,4,2 &$-1.57598 - 0.192268\,M - 0.00103889\,M^2 + 0.0000241287\,M^3$\\
0,5,1 &$-0.517589 - 0.178763\,M - 0.0115596\,M^2 + 0.000202814\,M^3 + 
     3.35858 \times {10}^{-6}\,M^4$\\
0,6,0 &$-0.0575099 - 0.0486174\,M - 0.0112156\,M^2 - 0.000619664\,M^3 + 
     0.0000116406\,M^4 + 1.86588 \times {10}^{-7}\,M^5$\\
1,0,5 &$-0.970081$\\
1,1,4 &$-4.81696 - 0.0334495\,M$\\
1,2,3 &$-9.40257 - 0.29895\,M + 0.00070522\,M^2$\\
1,3,2 &$-8.70366 - 0.994183\,M - 0.00319474\,M^2 + 0.00022649\,M^3$\\
1,4,1 &$-3.57656 - 1.20315\,M - 0.0721365\,M^2 + 0.00139263\,M^3 + 
     0.0000411225\,M^4$\\
1,5,0 &$-0.476875 - 0.398857\,M - 0.0898279\,M^2 - 0.00462342\,M^3 + 
     0.000098325\,M^4 + 2.7415 \times {10}^{-6}\,M^5$\\
2,0,4 &$-3.4864$\\
2,1,3 &$-13.6166 - 0.329002\,M$\\
2,2,2 &$-18.9348 - 1.97645\,M - 0.00712793\,M^2$\\
2,3,1 &$-10.382 - 3.38062\,M - 0.185104\,M^2 + 0.00215554\,M^3$\\
2,4,0 &$-1.73034 - 1.4286\,M - 0.312569\,M^2 - 0.0150661\,M^3 + 
     0.000179628\,M^4$\\
3,0,3 &$-7.02194$\\
3,1,2 &$-19.5576 - 1.50824\,M$\\
3,2,1 &$-16.0889 - 4.7868\,M - 0.190108\,M^2$\\
3,3,0 &$-3.57531 - 2.85139\,M - 0.574113\,M^2 - 0.0211231\,M^3$\\
4,0,2 &$-8.19093$\\
4,1,1 &$-13.4675 - 2.91439\,M$\\
4,2,0 &$-4.48915 - 3.21604\,M - 0.485732\,M^2$\\
5,0,1 &$-4.86262$\\
5,1,0 &$-3.24175 - 1.62087\,M$\\
6,0,0 &$-1.03493$\\
\end{tabular}
\end{table}

\begin{table}[tbp]
\vspace{0truecm}
\renewcommand\arraystretch{0.65}
\squeezetable
\caption{
The coefficients $b^{(v)}_{ijk}$, cf. Eq. (\ref{bv}).
}
\label{betavc}
\begin{tabular}{cl}
\multicolumn{1}{c}{$i,j,k$}&
\multicolumn{1}{c}{$R_M^{-j-1} b^{(v)}_{ijk}$}\\
\tableline \hline

0,0,3 &$0.0803991 + 0.0100499\,M$\\
0,1,2 &$0.369295 + 0.0484367\,M + 0.000284357\,M^2$\\
0,2,1 &$0.602154 + 0.0995834\,M + 0.00303927\,M^2$\\
0,3,0 &$0.243427 + 0.0974186\,M + 0.0100186\,M^2 + 0.0002056\,M^3$\\
1,0,2 &$0.564685 + 0.0705856\,M$\\
1,1,1 &$1.80131 + 0.266837\,M + 0.00520921\,M^2$\\
1,2,0 &$1.0441 + 0.397331\,M + 0.0362666\,M^2 + 0.000364277\,M^3$\\
2,0,1 &$1.44157 + 0.180197\,M$\\
2,1,0 &$1.56695 + 0.524919\,M + 0.0411313\,M^2$\\
3,0,0 &$0.814816 + 0.101852\,M$\\
\hline
0,0,4 &$-0.0670637 - 0.00838297\,M$\\
0,1,3 &$-0.376283 - 0.0486096\,M - 0.000196771\,M^2$\\
0,2,2 &$-0.857759 - 0.118296\,M - 0.00138446\,M^2$\\
0,3,1 &$-0.838966 - 0.178873\,M - 0.00870044\,M^2 + 0.0000687335\,M^3$\\
0,4,0 &$-0.248229 - 0.112659\,M - 0.0150567\,M^2 - 0.000585531\,M^3 + 
     2.63588 \times {10}^{-6}\,M^4$\\
1,0,3 &$-0.548393 - 0.0685491\,M$\\
1,1,2 &$-2.46028 - 0.326888\,M - 0.00241906\,M^2$\\
1,2,1 &$-3.52974 - 0.717174\,M - 0.0328982\,M^2 + 0.000199551\,M^3$\\
1,3,0 &$-1.35535 - 0.598611\,M - 0.0763386\,M^2 - 0.00279075\,M^3 + 
     5.68096 \times {10}^{-6}\,M^4$\\
2,0,2 &$-1.86702 - 0.233378\,M$\\
2,1,1 &$-5.18827 - 0.965409\,M - 0.0396094\,M^2$\\
2,2,0 &$-2.87884 - 1.21126\,M - 0.142369\,M^2 - 0.0044929\,M^3$\\
3,0,1 &$-2.67974 - 0.334967\,M$\\
3,1,0 &$-2.83487 - 0.985509\,M - 0.0788938\,M^2$\\
4,0,0 &$-1.09216 - 0.136521\,M$\\
\hline
0,0,5 &$0.0777718 + 0.00972148\,M$\\
0,1,4 &$0.498091 + 0.0634586\,M + 0.000149661\,M^2$\\
0,2,3 &$1.36139 + 0.179665\,M + 0.00118632\,M^2$\\
0,3,2 &$1.96208 + 0.303336\,M + 0.00757647\,M^2 + 0.0000396103\,M^3$\\
0,4,1 &$1.35191 + 0.341883\,M + 0.0225965\,M^2 + 0.000151344\,M^3 + 
     3.53109 \times {10}^{-6}\,M^4$\\
0,5,0 &$0.312662 + 0.156794\,M + 0.0247779\,M^2 + 0.00130674\,M^3 + 
     6.86391 \times {10}^{-6}\,M^4 + 9.64099 \times {10}^{-8}\,M^5$\\
1,0,4 &$0.685931 + 0.0857414\,M$\\
1,1,3 &$3.73862 + 0.480512\,M + 0.00164801\,M^2$\\
1,2,2 &$8.02393 + 1.1954\,M + 0.0248927\,M^2 + 0.000105193\,M^3$\\
1,3,1 &$7.26959 + 1.78276\,M + 0.110756\,M^2 + 0.000298431\,M^3 + 
     0.000013906\,M^4$\\
1,4,0 &$2.06581 + 1.01732\,M + 0.155716\,M^2 + 0.00765683\,M^3 + 
     8.67851\,{10}^{-6}\,M^4 + 2.53675 \times {10}^{-7}\,M^5$\\
2,0,3 &$2.68432 + 0.33554\,M$\\
2,1,2 &$11.3782 + 1.62939\,M + 0.0258891\,M^2$\\
2,2,1 &$15.1532 + 3.58071\,M + 0.208316\,M^2 - 0.000313018\,M^3$\\
2,3,0 &$5.60954 + 2.7006\,M + 0.397804\,M^2 + 0.0180874\,M^3 - 
     0.0000496743\,M^4$\\
3,0,2 &$5.6561 + 0.707013\,M$\\
3,1,1 &$14.6374 + 3.06149\,M + 0.153977\,M^2$\\
3,2,0 &$7.87876 + 3.54347\,M + 0.468803\,M^2 + 0.018622\,M^3$\\
4,0,1 &$5.56346 + 0.695432\,M$\\
4,1,0 &$5.75768 + 2.06202\,M + 0.167789\,M^2$\\
5,0,0 &$1.75644 + 0.219555\,M$\\
\hline
0,0,6 &$-0.104845 - 0.0131056\,M$\\
0,1,5 &$-0.766484 - 0.0974606\,M - 0.000206259\,M^2$\\
0,2,4 &$-2.44861 - 0.314595\,M - 0.00106485\,M^2$\\
0,3,3 &$-4.38376 - 0.608037\,M - 0.00748317\,M^2 + 3.14201 \times {10}^{-6}\,M^3$\\
0,4,2 &$-4.54538 - 0.806121\,M - 0.0303046\,M^2 - 0.0000528742\,M^3 + 
     2.15704 \times {10}^{-6}\,M^4$\\
0,5,1 &$-2.39859 - 0.70043\,M - 0.0596949\,M^2 - 0.00115086\,M^3 + 
     8.26434 \times {10}^{-6}\,M^4 + 2.2795 \times {10}^{-7}\,M^5$\\
0,6,0 &$-0.453777 - 0.247702\,M - 0.0450631\,M^2 - 0.00309346\,M^3 - 
     0.000053787\,M^4 +$ \\
&$+2.63266 \times {10}^{-7}\,M^5 + 4.89741 \times {10}^{-9}\,M^6$\\
1,0,5 &$-1.0206 - 0.127575\,M$\\
1,1,4 &$-6.51597 - 0.822062\,M - 0.000945689\,M^2$\\
1,2,3 &$-17.4536 - 2.36625\,M - 0.0229946\,M^2 + 9.24619 \times {10}^{-6}\,M^3$\\
1,3,2 &$-23.9918 - 4.14517\,M - 0.145391\,M^2 - 0.000191685\,M^3 + 
     9.11727 \times {10}^{-6}\,M^4$\\
1,4,1 &$-15.6729 - 4.48444\,M - 0.369036\,M^2 - 0.00652617\,M^3 + 
     0.000027074\,M^4 + 1.11884 \times {10}^{-6}\,M^5$\\
1,5,0 &$-3.51598 - 1.89485\,M - 0.337579\,M^2 - 0.0223394\,M^3 - 
     0.000357755\,M^4 + $\\
&$+4.31553 \times {10}^{-7}\,M^5 + 1.52437\times {10}^{-8}\,M^6$\\
2,0,4 &$-4.4801 - 0.560012\,M$\\
2,1,3 &$-23.8866 - 3.18735\,M - 0.0251901\,M^2$\\
2,2,2 &$-48.8238 - 8.24755\,M - 0.271114\,M^2 - 0.000380338\,M^3$\\
2,3,1 &$-41.9591 - 11.7782\,M - 0.936933\,M^2 - 0.0151248\,M^3 - 
     0.0000114803\,M^4$\\
2,4,0 &$-11.5837 - 6.16458\,M - 1.07549\,M^2 - 0.0685196\,M^3 - 
     0.000997932\,M^4 - 3.16906 \times {10}^{-6}\,M^5$\\
3,0,3 &$-11.3503 - 1.41879\,M$\\
3,1,2 &$-45.7836 - 7.2337\,M - 0.188844\,M^2$\\
3,2,1 &$-57.9214 - 15.4564\,M - 1.13636\,M^2 - 0.0136677\,M^3$\\
3,3,0 &$-20.8832 - 10.7783\,M - 1.79165\,M^2 - 0.105048\,M^3 - 0.00108945\,M^4
    $\\
4,0,2 &$-16.8465 - 2.10582\,M$\\
4,1,1 &$-41.5421 - 9.44192\,M - 0.531144\,M^2$\\
4,2,0 &$-21.8649 - 10.3417\,M - 1.48269\,M^2 - 0.066452\,M^3$\\
5,0,1 &$-12.4665 - 1.55831\,M$\\
5,1,0 &$-12.683 - 4.64788\,M - 0.382814\,M^2$\\
6,0,0 &$-3.1978 - 0.399725\,M$\\
\end{tabular}
\end{table}

\begin{table}[tbp]
\vspace{0truecm}
\renewcommand\arraystretch{0.7}
\squeezetable
\caption{
The coefficients $b^{(w)}_{ijk}$, cf. Eq. (\ref{bw}).
}
\label{betawc}
\begin{tabular}{cl}
\multicolumn{1}{c}{$i,j,k$}&
\multicolumn{1}{c}{$R_M^{-j} b^{(w)}_{ijk}$}\\
\tableline \hline
0,0,3 &$0.35107$\\
0,1,2 &$1.31383$\\
0,2,1 &$1.68533 + 0.00307141\,M$\\
0,3,0 &$0.643805 + 0.0574128\,M - 0.0017162\,M^2$\\
1,0,2 &$1.47806$\\
1,1,1 &$3.79199 + 0.00691068\,M$\\
1,2,0 &$2.17284 + 0.193768\,M - 0.00579216\,M^2$\\
2,0,1 &$2.133$\\
2,1,0 &$2.44445 + 0.199774\,M$\\
3,0,0 &$0.916668$\\
\hline
0,0,4 &$-0.376527$\\
0,1,3 &$-1.80719$\\
0,2,2 &$-3.34772 + 0.00754184\,M$\\
0,3,1 &$-2.73858 - 0.0492189\,M - 0.0000262347\,M^2$\\
0,4,0 &$-0.767062 - 0.0890547\,M + 0.0000407114\,M^2 - 0.0000875861\,M^3$\\
1,0,3 &$-2.03309$\\
1,1,2 &$-7.53237 + 0.0169691\,M$\\
1,2,1 &$-9.24272 - 0.166114\,M - 0.0000885421\,M^2$\\
1,3,0 &$-3.45178 - 0.400746\,M + 0.000183201\,M^2 - 0.000394138\,M^3$\\
2,0,2 &$-4.23696$\\
2,1,1 &$-10.3981 - 0.212414\,M$\\
2,2,0 &$-5.82488 - 0.690232\,M - 0.00248026\,M^2$\\
3,0,1 &$-3.89927$\\
3,1,0 &$-4.36866 - 0.415187\,M$\\
4,0,0 &$-1.22868$\\
\hline
0,0,5 &$0.495548$\\
0,1,4 &$2.88579$\\
0,2,3 &$6.8946 - 0.0230874\,M$\\
0,3,2 &$8.36453 + 0.00396206\,M + 0.000213631\,M^2$\\
0,4,1 &$4.98655 + 0.175728\,M - 0.00207184\,M^2 - 0.0000193829\,M^3$\\
0,5,0 &$1.09653 + 0.157913\,M + 0.00235846\,M^2 - 0.0000614713\,M^3 - 
     5.38712 \times {10}^{-6}\,M^4$\\
1,0,4 &$3.24652$\\
1,1,3 &$15.5129 - 0.0519466\,M$\\
1,2,2 &$28.2303 + 0.0133719\,M + 0.000721005\,M^2$\\
1,3,1 &$22.4395 + 0.790776\,M - 0.00932327\,M^2 - 0.0000872231\,M^3$\\
1,4,0 &$6.16801 + 0.88826\,M + 0.0132664\,M^2 - 0.000345776\,M^3 - 
     0.0000303026\,M^4$\\
2,0,3 &$8.72598$\\
2,1,2 &$31.7591 + 0.124994\,M$\\
2,2,1 &$37.8666 + 1.47045\,M - 0.0146658\,M^2$\\
2,3,0 &$13.878 + 2.04994\,M + 0.0348703\,M^2 - 0.000849221\,M^3$\\
3,0,2 &$11.9097$\\
3,1,1 &$28.4 + 0.965884\,M$\\
3,2,0 &$15.6128 + 2.22247\,M + 0.0408606\,M^2$\\
4,0,1 &$7.98749$\\
4,1,0 &$8.78218 + 0.933899\,M$\\
5,0,0 &$1.97599$\\
\hline
0,0,6 &$-0.749689$\\
0,1,5 &$-5.12987$\\
0,2,4 &$-14.928 + 0.0486813\,M$\\
0,3,3 &$-23.5697 + 0.0957169\,M - 0.00083904\,M^2$\\
0,4,2 &$-21.0735 - 0.166287\,M - 0.0000148277\,M^2 + 4.49885 \times {10}^{-6}\,M^3$
    \\
0,5,1 &$-9.82983 - 0.53385\,M + 0.00220333\,M^2 - 0.000130668\,M^3 - 
     2.59594 \times {10}^{-6}\,M^4$\\
0,6,0 &$-1.77455 - 0.304043\,M - 0.00943381\,M^2 + 0.0000669939\,M^3 - 
     6.57249 \times {10}^{-6}\,M^4 - 3.75311 \times {10}^{-7}\,M^5$\\
1,0,5 &$-5.77111$\\
1,1,4 &$-33.588 + 0.109533\,M$\\
1,2,3 &$-79.5478 + 0.323044\,M - 0.00283176\,M^2$\\
1,3,2 &$-94.8309 - 0.748291\,M - 0.0000667246\,M^2 + 0.0000202448\,M^3$\\
1,4,1 &$-55.2928 - 3.0029\,M + 0.0123937\,M^2 - 0.000735009\,M^3 - 
     0.0000146022\,M^4$\\
1,5,0 &$-11.9782 - 2.05229\,M - 0.0636782\,M^2 + 0.000452209\,M^3 - 
     0.0000443643\,M^4 - 2.53335 \times {10}^{-6}\,M^5$\\
2,0,4 &$-18.8932$\\
2,1,3 &$-89.4913 + 0.0651677\,M$\\
2,2,2 &$-160.027 - 1.81428\,M + 0.00813333\,M^2$\\
2,3,1 &$-124.409 - 7.20262\,M + 0.023168\,M^2 - 0.00136807\,M^3$\\
2,4,0 &$-33.6888 - 5.89631\,M - 0.193047\,M^2 + 0.00140166\,M^3 - 
     0.000108167\,M^4$\\
3,0,3 &$-33.5592$\\
3,1,2 &$-120.02 - 1.44746\,M$\\
3,2,1 &$-139.96 - 8.18899\,M - 0.000367593\,M^2$\\
3,3,0 &$-50.5332 - 8.87281\,M - 0.304904\,M^2 + 0.00178302\,M^3$\\
4,0,2 &$-33.7557$\\
4,1,1 &$-78.7274 - 3.5874\,M$\\
4,2,0 &$-42.6374 - 6.91385\,M - 0.200772\,M^2$\\
5,0,1 &$-17.7137$\\
5,1,0 &$-19.1868 - 2.21571\,M$\\
6,0,0 &$-3.59753$\\
\end{tabular}
\end{table}

% ========================= REFERENCES =========================


\begin{references}
\bibitem{tetra}
We mention that the Hamiltonian (\ref{Hphi4}) with $M=2$ and
$N\geq 2$ is physically interesting because it
describes the critical properties in some structural and
antiferromagnetic phase transitions, 
see, e.g., Refs.~\cite{Mukamel-75,Aharony-76}. 
For recent studies, see, e.g., Refs.~\cite{PV-r,MV-01,SV-99} and 
references therein.
\bibitem{footnote4}
For comparison, we have also performed the direct analysis of  
$\omega_v$, resumming the expansions for fixed 
$\wb/\ub$. In zero dimensions, these series are not Borel summable, 
and this is expected to be true in any dimension. However, for the 
short series we are considering, we can still hope to obtain 
reasonable results. 
We obtain $\omega_v=-0.023(20)\{20\}$ (the second error is related
to the uncertainty on the coordinates of the fixed point)
in agreement with the one given in the text.
\bibitem{extc}
We are indebted with Christian Tanguy for informing us 
of the systematic relation of the coefficients
of the series for the cubic model 
that corresponds to Eq.~(\ref{euevrel}).
\end{references}
\end{document}